\documentclass[a4paper,pra,twocolumn,preprintnumbers,showpacs]{revtex4}\newif\ifnarrow\narrowtrue

\makeatletter\def\Dated@name{}\makeatother

\usepackage{amsmath,amssymb,amsfonts}
\usepackage{bm,graphicx,color,natbib}

\definecolor{highlight}{rgb}{0,0,1}

\newcommand{\TR}{\text{Tr}}
\newcommand{\mydagger}{{\dagger}}
\newcommand{\phdagger}{{\phantom{\mydagger}\!}}
\newcommand{\ETAL}{{\em et al.}}
\newcommand{\bra}[1]{\langle{#1}|}
\newcommand{\ket}[1]{|{#1}\rangle}
\newcommand{\braket}[2]{\langle{#1}|{#2}\rangle}
\newcommand{\expval}[1]{\langle{#1}\rangle}
\newcommand{\calI}{{\cal I}}
\newcommand{\calK}{{\cal K}}
\newcommand{\mycasea}{(a)}
\newcommand{\mycaseb}{(b)}
\newcommand{\OBS}{O}
\newcommand\myamp{&}

\begin{document}

  \title{Relaxation of a one-dimensional Mott insulator after an interaction quench}

  \author{Marcus Kollar and Martin Eckstein}

  \affiliation{Theoretical Physics III, Center for Electronic
    Correlations and Magnetism, Institute of Physics, University of
    Augsburg, 86135 Augsburg, Germany}
  
  \date{April 28, 2008}

  \begin{abstract}
    \vspace*{6mm}%
    We obtain the exact time evolution for the one-dimensional
    integrable fermionic $1/r$ Hubbard model after a sudden change of
    its interaction parameter, starting from either a metallic or a
    Mott-insulating eigenstate.  In all cases the system relaxes to a
    new steady state, showing that the presence of the Mott gap does
    not inhibit relaxation.  The properties of the final state are
    described by a generalized Gibbs ensemble.  We discuss under which
    conditions such ensembles provide the correct statistical
    description of isolated integrable systems in general.  We find
    that generalized Gibbs ensembles do predict the properties of 
    the steady state correctly, provided that the observables or 
    initial states are sufficiently uncorrelated in terms of the 
    constants of motion.
  \end{abstract}

  \pacs{03.75.Ss, 05.30.Fk, 71.27.+a, 02.30.Ik%
    \vspace*{6mm}%
  }

  \maketitle

  \section{Introduction}
  
  Recent experiments with ultracold atomic
  gases~\cite{Bloch2007pre,Greiner2002b,Kinoshita2006,Hofferberth07a}
  have made it possible to study the time evolution of a tunable quantum
  many-body system that is kept in excellent isolation from the
  environment.  For example, such a quantum system can be forced out
  of equilibrium by suddenly changing a parameter in the Hamiltonian.
  Then the system may or may not relax to a new steady state, which is
  not necessarily the thermal state predicted by statistical
  mechanics.  After such a ``quantum quench'' the system evolves
  according to Schr{\"o}dinger's equation
  \begin{align}
    \ket{\Psi(t)}
    =
    \exp(-iHt/\hbar)
    \ket{\Psi(0)}
    \,,\label{eq:psi_of_t}
  \end{align}
  where $\ket{\Psi(0)}$ is the prepared initial state and $H$ is the
  new Hamiltonian for times $t$ $\geq$ $0$. This situation has
  recently been studied by a variety of numerical and analytical
  techniques~\cite{Altman2002a,Polkovnikov2002a,%
    Sengupta04,Cazalilla06,Rigol06+07,%
    Kollath07,Manmana07,Cramer2008,Eckstein07,%
    Rigol2008a,Gangardt2008,Barthel2008,%
    Moeckel08a,Anders2005}.

  Due to the unitary time evolution the wave function $\ket{\Psi(t)}$
  of an isolated system remains pure for all times and does not
  converge for $t$ $\to$ $\infty$. Only the state of a finite
  subsystem, for which the rest of the system effectively acts as a
  reservoir, can become stationary~\cite{Cramer2008,Barthel2008}. Nevertheless,
  also for the entire system we expect \emph{relaxation} of the
  expectation value $\bra{\Psi(t)}\OBS\ket{\Psi(t)}$ of an observable
  $\OBS$ to a stationary value for large times.  However, this global
  relaxation can happen only for (i)~sufficiently large systems,
  (ii)~sufficiently simple observables, and (iii) sufficiently
  complicated Hamiltonians, for the following reasons.  First of all,
  (i)~many degrees of freedom are needed, so that the thermodynamic
  limit may be taken, otherwise one expects finite recurrence
  times~\cite{Montroll1961} (see~\cite{Rigol2005a} for a recent
  example).  Furthermore, (ii) the expectation value of a complicated
  observable need not relax; for example, the expectation value of
  $\OBS$ $=$ $\ket{n_1}\bra{n_1}$ $+$ $\ket{n_2}\bra{n_2}$, involving
  the projectors onto two eigenstates of $H$ with different energies,
  oscillates for all times.  Usually such projectors are highly
  nonlocal and their expectation values correspond to correlation
  functions of very high order. On the other hand, local and
  few-particle observables are usually simple enough to relax to new
  stationary values. Finally, (iii) the Hamiltonian $H$ that governs
  the dynamics must also be sufficiently complicated.  For example,
  the magnetization of an Ising chain in a transverse magnetic field
  relaxes for long-range interactions, but keeps oscillating when only
  next neighbors are coupled~\cite{Heims65}.  Similar
  ``collapse-and-revival'' oscillations of a many-body system were
  recently observed in experiments with ultracold atoms by Greiner
  \ETAL{}~\cite{Greiner2002b}. In their experiments, a Bose condensate
  was prepared in the potential of an optical lattice which was
  suddenly steepened~\cite{Greiner2002b}, then the bosons are
  essentially only subject to the Hubbard interaction $H$ $=$
  $U\sum_in_i(n_i-1)$ but no hopping between lattice sites occurs.
  Since in this case $H/U$ has only integer eigenvalues, the wave
  function will oscillate for all times with period $2\pi U/\hbar$.

  For small hopping between lattice sites it follows from perturbation
  theory that expectation values keep oscillating for short times.
  But it is not clear what will happen for long observation times.  Is
  relaxation possible for a bosonic or fermionic Hubbard model if the
  spectrum has a Mott gap, or if the initial state is a Mott
  insulator? Or is it prevented by the Mott gap in the energy
  spectrum?  The answer is no, not necessarily: in
  Sec.~\ref{sec:hubbardmodel} we provide an example, the $1/r$
  fermionic Hubbard model~\cite{Gebhard92}, which shows that
  relaxation in the presence of a Mott gap is indeed possible.
  Note that the formation of a fermionic Mott insulator was recently
  observed with ultracold atoms~\cite{Jordens2008pre}.

  Another central question is whether the steady state of a quenched
  isolated system can be described by an effective density
  matrix~$\rho$, such that $\TR[\OBS\rho]$ yields the correct
  expectation value for any observable $\OBS$ which relaxes.
  Statistical mechanics can be used to make an approximate but usually
  accurate prediction $\rho_{\text{mic}}$ for this steady-state
  density matrix.  For example, the microcanonical prediction is that
  $\rho_{\text{mic}}$ $=$ $\text{const}$ for states with energy close
  to $\bra{\Psi(0)}H\ket{\Psi(0)}$, and zero otherwise. If
  $\TR[\OBS\rho_{\text{mic}}]$ indeed agrees with the long-time limit
  of $\bra{\Psi(t)}\OBS\ket{\Psi(t)}$, we say that the system
  \emph{thermalizes}. Clearly, thermalization can be expected only for
  sufficiently coarse-grained observables; it is always possible to
  construct a complicated correlation function that depends on the
  details of the initial conditions and is not described by
  $\rho_{\text{mic}}$.  As for classical gases, thermalization is
  generally expected for isolated interacting quantum systems.
  Indeed, for one-dimensional atomic Bose gases, the dynamics leading
  to the thermal state were recently observed by Hofferberth
  \ETAL{}~\cite{Hofferberth07a}.
 
  Thermalization of an isolated system is impossible in certain cases,
  usually because the system is integrable in the sense that there are
  infinitely many constants of motion.  Simple theoretical examples
  are one-dimensional integrable models, such as the $XY$ spin chain,
  for which the magnetization does not thermalize after a quench of
  the longitudinal magnetic
  field~\cite{Girardeau1969,Barouch1969,Girardeau1970,Barouch1971},
  or the Ising
  chain in a transverse field, whose correlation functions do not
  thermalize after a field quench~\cite{Sengupta04}.  Experimentally,
  the lack of thermalization was recently observed by Kinoshita
  \ETAL{}~\cite{Kinoshita2006} for bosonic atoms confined to one
  dimension, whose nonthermal stationary momentum distributions were
  attributed to integrability.

  The fundamental assumption of statistical mechanics is that the
  equilibrium state is characterized only by a few thermodynamic
  variables such as internal energy and particle number. However if
  the system is integrable, its infinitely many constants of motion
  lead to a rather detailed memory of the initial state, because much
  fewer states are accessible to the dynamics.  Nevertheless, even in
  the absence of thermalization, a statistical prediction for the
  steady state can be made with a \emph{generalized Gibbs
    ensemble~(GGE)}, as discussed in the recent work of Rigol
  \ETAL{}~\cite{Rigol06+07}.  For example, if the time evolution
  of an integrable system is determined by the effective Hamiltonian
  \begin{align}
    H_{\text{eff}}
    &=
    \sum_\alpha\epsilon_\alpha\;\calI_\alpha
    \,,\label{eq:effH}
  \end{align}
  where the operators $\calI_\alpha$ commute,
  ${[\calI_\alpha,\calI_\beta]}$ $=$ $0$, then the standard choice for
  the statistical operator of the GGE is constructed from these
  constants of motion according to~\cite{Rigol06+07}%
  \begin{align}%
    \rho_{\text{G}}^\phdagger
    &=
    \frac{e^{-\sum_\alpha\lambda_\alpha\calI_\alpha}}{Z_{\text{G}}^\phdagger}
    \,,
    &
    Z_{\text{G}}^\phdagger
    &=
    \TR[e^{-\sum_\alpha\lambda_\alpha\calI_\alpha}]
    \,,\label{eq:gge}%
  \end{align}%
  This choice maximizes the entropy ($S$ $=$ $-\,\TR[\rho\ln\rho]$)
  for fixed expectation values
  $\expval{\calI_\alpha}_\text{G}^\phdagger$, which are set to their
  initial-state expectation values
  $\expval{\calI_\alpha}_{0}^\phdagger$ by an appropriate choice for
  the Lagrange multipliers $\lambda_\alpha$~\cite{Balian91a}.  The GGE
  prediction for the steady-state expectation value of an observable
  $\OBS$ is then $\expval{\OBS}_\text{G}^\phdagger$ $=$
  $\TR[\OBS\rho_{\text{G}}^\phdagger]$.  GGEs successfully predict
  some properties of the nonthermal steady states occurring after
  quenches in integrable or highly constrained systems.  For example,
  they yield the correct nonthermal momentum distribution of
  one-dimensional hard-core bosons~\cite{Rigol06+07}
  (experimentally realized in Ref.~\onlinecite{Kinoshita2006}), for
  the one-dimensional Tomonaga-Luttinger model~\cite{Cazalilla06}, and
  for the Falicov-Kimball model in infinite
  dimensions~\cite{Eckstein07}.
  The GGE also yields the correct double occupation for the
  $1/r$ fermionic Hubbard model, as shown in
  Sec.~\ref{sec:hubbardmodel} below.

  However, in some cases, the stationary values of some observables
  differ from the statistical predictions of the GGE. For
  one-dimensional hard-core bosons the unit-cell averaged one-particle
  correlation function is not described by the
  GGE~\cite{Rigol06+07}.  Furthermore Gangardt and
  Pustilnik~\cite{Gangardt2008} pointed out that the
  GGE~(\ref{eq:gge}) may not capture correlations between the
  conserved quantities $\calI_\alpha$.  As a consequence the merit of
  generalized Gibbs ensembles is currently somewhat controversial.  In
  this situation rather general criteria for the validity of GGEs
  should be useful, which we derive in
  Sec.~\ref{sec:gibbsensembles}.  Our approach is complementary to the
  work of Barthel and Schollw{\"o}ck~\cite{Barthel2008}, who recently 
  showed that for finite subsystems the reduced density matrix converges
  to the GGE under certain mathematical conditions on the initial
  state and Hamiltonian.

  It should be noted that an important ambiguity lingers in the
  construction of the GGE for the Hamiltonian~(\ref{eq:effH}).  While
  all the $\calI_\alpha$ are conserved, so are all their combinations,
  i.e., all products of the form $\calI_\alpha\calI_\beta$,
  $\calI_\alpha\calI_\beta\calI_\gamma$,~\ldots, leading to the
  question of whether all such products should also be included in the
  exponent of the density matrix~(\ref{eq:gge}).  In
  Sec.~\ref{sec:gibbsensembles} we show that for observables and
  initial states which involve little or no correlation, it suffices to
  fix the constraints $\calI_\alpha$, as in Eq.~(\ref{eq:gge}), but
  not their products.  We also provide an example for which different
  choices of the GGE lead to different predictions.

  Finally, we note that thermalization might also be prevented in
  nonintegrable systems due to many-body effects, e.g., the presence
  of a Mott gap.  Nonthermal steady states in nonintegrable systems
  were observed and argued for in recent numerical studies for finite
  one-dimensional soft-core bosons~\cite{Kollath07} and spinless
  fermions~\cite{Manmana07}.  By contrast, thermalization was observed
  for hard-core bosons on a two-dimensional lattice~\cite{Rigol2008a}.
  Fast relaxation to a nonthermal quasisteady state, so-called
  prethermalization~\cite{Berges2004a}, was recently observed for the
  fermionic Hubbard model in high dimensions~\cite{Moeckel08a}.
  Further studies of relaxation in nonintegrable many-body systems
  are therefore desirable, but will not be the subject of this paper.

  Our goals in the present paper are thus two-fold. (i) On the one
  hand, we provide an explicit example of relaxation in a fermionic
  Mott insulator: In Sec.~\ref{sec:hubbardmodel} we obtain the
  exact time evolution of the one-dimensional fermionic Hubbard model
  with $1/r$ hopping and repulsive interaction $U$~\cite{Gebhard92},
  starting from the metallic ground state at $U$ $=$ $0$ or the
  insulating ground state at $U$ $=$ $\infty$. We find that the
  expectation value of the double occupation $d(t)$ relaxes with
  algebraically damped oscillations to a new stationary value for all
  $U$, i.e., relaxation is not inhibited by the presence of a Mott
  gap. The long-time limit $d_\infty$ $=$ $\lim_{t\to\infty}d(t)$
  differs from the thermal value, as expected for an integrable
  system, but is described by an appropriate GGE.  (ii) On the other
  hand we discuss under which circumstances GGEs describe the steady
  state of integrable systems after relaxation.  We show in
  Sec.~\ref{sec:gibbsensembles} that their validity depends on the
  observable, on correlations in the initial state, and possibly the
  system size.


  \section{Interaction quench in a fermionic Hubbard model}\label{sec:hubbardmodel}

  \subsection*{$1/r$ Hubbard chain}
  
  We consider sudden changes in the interaction parameter of the
  one-dimensional $1/r$ fermionic Hubbard model,
  \begin{subequations}%
    \begin{align}%
      H
      &=
      H_0
      +
      H_1
      \,,
      ~~~~~~
      H_1
      =
      U
      \sum_i
      n_{i\uparrow}
      n_{i\downarrow}
      \,,
      \\
      H_0
      &=
      \sum_{\substack{i,j=1..L\\\sigma=\uparrow,\downarrow}} \! t_{ij}\,
      c_{i\sigma}^\mydagger
      c_{j\sigma}^\phdagger
      =
      \sum_{\substack{|k|<\pi\\\sigma=\uparrow,\downarrow}} \! \epsilon_k\,
      c_{k\sigma}^\mydagger
      c_{k\sigma}^\phdagger
      \,,
    \end{align}\label{eq:GR-model}%
  \end{subequations}%
  with repulsive on-site interaction $U$, bandwidth $W$, and
  dispersion $\epsilon_k$=$Wk/(2\pi)$, which corresponds to hopping
  amplitudes  $t_{mj}$ $=$ $(-iW/2L)(-1)^{m-j}/\sin[\pi(m-j)/L]$
  that decay proportionally to inverse distance.  This
  lattice model was introduced by Gebhard and
  Ruckenstein~\cite{Gebhard92} as a parent system of the $1/r^2$
  Haldane-Shastry Heisenberg chain~\cite{Haldane88,Shastry88}, to
  which it reduces in the limit of large $U$ for a half-filled band
  with density $n$ $=$ $1$.  We consider only $n$ $\leq$ $1$; larger
  densities can be treated by means of a particle-hole transformation,
  $c_{k\sigma}^\phdagger$ $\to$
  $c_{-k\sigma}^\mydagger$~\cite{Gebhard97}.  For $U$ $\geq$ $-W$ and
  any number of lattice sites $L$, the model~(\ref{eq:GR-model}) is 
  represented by an effective noninteracting bosonic Hamiltonian, from
  which ground-state and thermodynamic properties can be obtained
  analytically~\cite{Gebhard92,Gebhard94b,Gebhard97}.  For $U$ $=$ $0$
  the ground state of~(\ref{eq:GR-model}) is of course the Fermi sea,
  \begin{align}%
    \ket{\psi_0}
    &=
    \prod_{k<k_F}
    c_{k\uparrow}^\mydagger
    c_{k\downarrow}^\mydagger
    \;
    \ket{0}
    \,,
  \end{align}%
  with particle density $n$ $=$ $1$ $+$ $k_\text{F}/\pi$. For $U$ $=$
  $\infty$, on the other hand, the ground state
  is~\cite{Gebhard92,Gebhard94a}
  \begin{align}
    \ket{\psi_\infty}
    &=
    \prod_{i}
    (1-n_{i\uparrow}n_{i\downarrow})
    \;
    \ket{\psi_0}
    \,,
  \end{align}
  i.e., the Fermi sea with all doubly occupied sites projected out.
  At half-filling a Mott-Hubbard metal-insulator transition occurs at
  interaction strength $U_c=W$, with the Mott gap given by
  $\Delta=U-U_c$ for $U\geq U_c$~\cite{Gebhard92}. This
  metal-insulator transition is also captured by correlated
  variational wave functions~\cite{Gebhard94a,Dzierzawa95}.

  \subsection*{Interaction quenches}

  We now consider the following nonequilibrium situation. For times
  $t\leq0$ the system is prepared in the ground state for
  interaction parameter $U$ $=$ $0$ or $\infty$, i.e.,
  \begin{subequations}%
    \begin{align}%
      \ket{\Psi(0)}
      &=
      \frac{\ket{\psi_0}}{\sqrt{\braket{\psi_0}{\psi_0}}}
      &
      &\text{metallic state, or}\label{eq:inimetal}
      \\
      \ket{\Psi(0)}
      &=
      \frac{\ket{\psi_\infty}}{\sqrt{\braket{\psi_\infty}{\psi_\infty}}}
      &
      &\text{Mott insulator.}\label{eq:iniMott}
    \end{align}\label{eq:initialstates}%
  \end{subequations}%
  Then at time $t=0$ the interaction is suddenly switched to a new
  value $0<U<\infty$, so that the time evolution for $t\geq0$ is
  governed by the Hamiltonian~(\ref{eq:GR-model}), i.e., the system
  evolves according to Eq.~(\ref{eq:psi_of_t}).  We refer to these two
  types of quenches as $0$~$\to$~$U$ (starting from the metallic
  state~(\ref{eq:inimetal})) and $\infty$~$\to$~$U$ (starting from the
  Mott-insulating state~(\ref{eq:iniMott})), respectively.  More
  general initial states corresponding to intermediate values of $U$
  can also be used; they lead to similar results which are omitted
  here.

  \subsection*{Bosonic representation}

  In the bosonic representation of Ref.~\onlinecite{Gebhard94a} the
  initial states~(\ref{eq:initialstates}) factorize. They can be
  written in terms of hard-core bosons
  ($\bullet,\circ,\uparrow,\downarrow$) in the
  form~\cite{Gebhard92,Gebhard94a}%
  \begin{subequations}%
    \begin{align}%
      \ket{\psi_0}
      &=
      \Big|\;
      [\uparrow\,\downarrow]\cdots[\uparrow\,\downarrow]
      {}_{\phantom{0}\;}\Big|_{0\;}
      [\bullet\,\circ]\cdots[\bullet\,\circ]
      {}_{\phantom{{\calK}}}\Big|_{{\calK}_{\text{F}}}\!
      \circ\cdots\circ
      \Big\rangle
      \\
      &\equiv
      \prod_{\calK}{}'\;
      \ket{\psi_{{\calK},{\calK}+\Delta{\calK}}^0}^{\phdagger}
      \,,
      \\
      \ket{\psi_\infty}
      &=
      \prod_{\calK}{}'\,
      \Big(
      1-\bm{D}_{{\calK},{\calK}+\Delta{\calK}}
      \Big)~
      \ket{\psi_{{\calK},{\calK}+\Delta{\calK}}^0}^{\phdagger}
      \,,
    \end{align}%
  \end{subequations}%
  where $\Delta$ $=$ $2\pi/L$, ${\calK}_\text{F}$ $=$ $(2n-1)\pi$, and
  the prime indicates that only every other bosonic
  pseudo\-mo\-men\-tum ${\calK}$ $\in$ $(-\pi,\pi)$ appears. The
  Hamiltonian~(\ref{eq:GR-model}) only acts separately on each space
  spanned by the bracketed configurations $[\uparrow\,\downarrow]$
  $\equiv$ $\binom{1}{0}$ and $[\bullet\,\circ]$ $\equiv$
  $\binom{0}{1}$ for neighboring pseudo\-mo\-men\-ta, ${\calK}$ and
  ${\calK}+\Delta$~\cite{Gebhard92,Gebhard94a},%
  \begin{subequations}%
    \begin{align}%
      H_{\text{eff}}
      &=
      \sum_{\calK<{\calK}_\text{F}}\!\!\!{}'\,
      \Big(
      \bm{T}_{{\calK},{\calK}+\Delta{\calK}}
      +
      U
      \bm{D}_{{\calK},{\calK}+\Delta{\calK}}
      \Big)
      \,,
      \\
      \bm{T}_{{\calK},{\calK}+\Delta{\calK}}
      &=
      \frac{W}{2}
      \left(
        \begin{array}{cc}
          \text{sgn}({\calK})&0\\
          0&-\text{sgn}({\calK})
        \end{array}
      \right)
      ,
      \\
      \bm{D}_{{\calK},{\calK}+\Delta{\calK}}
      &=
      \frac{1}{2}
      \left(
        \begin{array}{cc}
          1-a_{\calK}&\sqrt{1-a_{\calK}^2}\\
          \sqrt{1-a_{\calK}^2}&1+a_{\calK}\\
        \end{array}
      \right)
      ,
    \end{align}\label{eq:GRM_Heff}%
  \end{subequations}%
  with $a_{\calK}$ $=$ $\text{sgn}({\calK})(2{\cal K}+\Delta)/(2\pi)$.
  In this representation it is then straightforward, although tedious,
  to obtain the propagator $\exp(-iHt)$, its action on
  $\ket{\Psi(0)}$, and the expectation value of the double occupation
  $d(t)$ $=$ $\bra{\Psi(t)}n_{i\uparrow}n_{i\downarrow}\ket{\Psi(t)}$.
  We take the thermodynamic limit, $L$ $\to$ $\infty$ with fixed
  density $n$; the sums over $\calK$ are then replaced by integrals
  which can be evaluated analytically.

  \subsection*{Results for the double occupation}

  Setting the bandwidth to $W$ $=$ $1$ (and also $\hbar$ $=$ 1), our
  results for the interaction quenches $0$~$\to$~$U$ and
  $\infty$~$\to$~$U$ can be written as
  \begin{subequations}%
    \begin{align}
      d(t)\Big|_{0\to U\phantom{\infty}\!\!\!}
      &=
      \, c_++f(t)
      \,,
      \\
      d(t)\Big|_{\infty\to U\phantom{0}\!\!\!}
      &=
      \frac{c_--f(t)}{U}
      \,,\label{eq:dynamics}
    \end{align}
  \end{subequations}
  for the two types of quenches, with the abbreviations
  \begin{align}
    c_\pm
    &=
    \frac{n^2}{8}
    \mp
    \frac{\Delta^2}{32U^2}
    \bigg[
    2nU
    +
    \Omega^2
    \ln\frac{\omega}{\Omega}
    \bigg]
    \,,\label{eq:longtimelimit}
    \\
    \!\!
    f(t)
    &=
    g(\Omega,t)
    -
    g(\omega,t)
    -\frac{n}{8}\frac{\omega t\sin(\omega t)+3\cos(\omega t)}{Ut^2}
    \,.\label{eq:ffunction}
  \end{align}
  These expressions involve several energy scales, apart from the
  interaction $U$ and the bandwidth $W$ ($=$ $1$), namely the Mott gap
  $\Delta$ $=$ $U-1$, the total bandwidth of the spectrum $\Omega$ $=$
  $2+\Delta$ $=$ $U+1$, and $\omega$ $=$ $\sqrt{\Omega^2-4Un}$, a
  characteristic density-dependent energy scale appearing in the holon
  and spinon excitation energies~\cite{Gebhard92,Gebhard94b}.  As
  functions of $U$ the constants $c_\pm$ have the remarkable symmetry
  that both are invariant under the replacement $U$ $\to$ $1/U$ (for
  all $n$).  The function $g(\eta,t)$ in Eq.~(\ref{eq:ffunction}) is
  given by
  \begin{align}
    g(\eta,t)
    =
    \frac{1}{32U^2}\Big[
    -\Omega^2\Delta^2\text{Ci}(\eta t)
    +\frac{6\!-\!\Delta^2t^2}{t^3}\eta\sin(\eta t)
    \ifnarrow\nonumber\\\fi
    +\frac{6\!-\!(\Delta^2\!+\!8U)t^2}{t^4}\cos(\eta t)
    \Big]
    \,,
  \end{align}
  where $\text{Ci}(x)=-\int_x^\infty\cos(y)/y$ is the integral cosine.
  For the quench to $U$ $=$ $U_c$ ($=$ $1$) this reduces to $c_\pm$
  $=$ $1/8$ and $f(t)=f_1(t)$, where
  \begin{align}    
    f_1(t)
    &=
    -\frac{2t^2+3}{16t^4}
    +\frac{3}{8t^3}\sin(2t)
    -\frac{4t^2-3}{16t^4}\cos(2t)
    \,.
  \end{align}

  We note that in all cases $d(t)$ relaxes with damped oscillations
  from its initial value $d(0)$ ($=$ $n^2/4$ for the metallic
  state~(\ref{eq:inimetal}) or 0 for the Mott
  insulator~(\ref{eq:iniMott})) to a new stationary value.  This
  long-time limit, $d_\infty$ $=$ $\lim_{t\to\infty}d(t)$, always
  exists, even when quenching to $U$ $=$ $0$, $U_c$, or $\infty$.
  For these final values of $U$ we find
  \begin{align}
    \lim_{t\to\infty}d(t)
    &=
    \begin{cases}
      \dfrac{n^2}{4}          &    U:~0\to\infty        
      \\[1.5ex]
      \dfrac{n^2(3-2n)}{6}    &    U:~\infty\to 0        
      \\[1.5ex]
      \dfrac{n^2}{8}          &    U:~0\to1\text{ or }\infty\to1
    \end{cases}
    \,.\label{eq:dinfty_results}
  \end{align}
  For the quench to $U$ $=$ $0$ we note in particular that the
  stationary value of the double occupation differs from the thermal
  value, $n^2/4$. This is discussed in detail at the end of
  Sec.~\ref{sec:gibbsensembles}.

  For general $U$ and $n$ the function~(\ref{eq:ffunction}) behaves
  asymptotically as
  \begin{align}
    f(t)
    =
    -\frac{n(1-n)}{2}
    \frac{\sin(\omega t)}{\omega t}
    -
    \frac{\cos(\Omega t)}{4Ut^2}
    \ifnarrow~~~~~~~~~~~~~~~~~~\nonumber\\\fi
    +
    \frac{(1-3n)\omega^2+(1-n)\Omega^2}{2\omega^2}
    \frac{\cos(\omega t)}{4Ut^2}
    +
    O\Big(\frac{1}{t^3}\Big)
  \end{align}
  for large times, which at half-filling reduces to perfect beating,
  \begin{align}
    c_\pm
    &=
    \frac{1}{8}
    \mp
    \bigg[
    \frac{(1-U)^2}{16U}
    +
    \frac{(1-U^2)^2}{16U^2}
    \ln\Big|\frac{1-U}{1+U}\Big|
    \bigg]
    \,,\label{eq:longtimelimit1}
    \\
    f(t)
    &=
    -\frac{\cos(\Omega t)+\cos(\Delta t)}{4Ut^2}    
    =
    -\frac{\cos(Ut)\cos(t)}{2Ut^2}    
    \,.
  \end{align}
  We conclude that the relaxation of $d(t)$ involves the frequencies
  $\omega$ and $\Omega$ and that it falls off rather slowly as $1/t$
  for densities $n$ $<$ $1$, or as $1/t^2$ for $n$ $=$ $1$. This type
  of algebraic decay is typical for one-dimensional
  systems~\cite{Calabrese06a}.  Mathematically it can be traced to the
  Riemann-Lebesgue lemma, i.e., the $O(1/t)$ decay of one-dimensional
  integrals over oscillating functions~\cite{Barthel2008,Olver1997}.

  \subsection*{Nonthermal steady states}

\begin{figure*}[p]%
    \centerline{%
      \hspace*{\fill}%
      \includegraphics[width=0.48\hsize,clip=true]{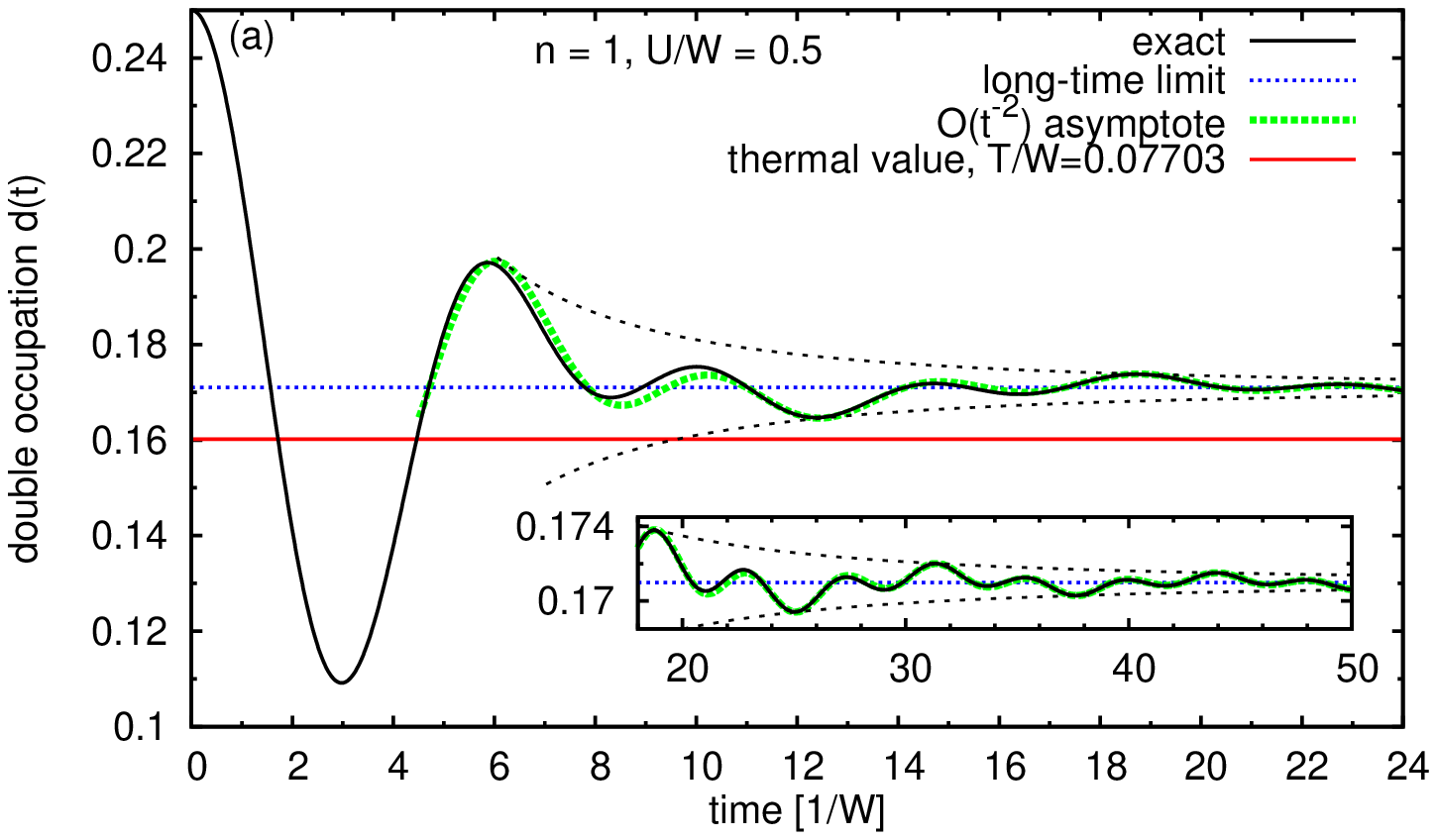}%
      \hspace*{\fill}%
      \includegraphics[width=0.48\hsize,clip=true]{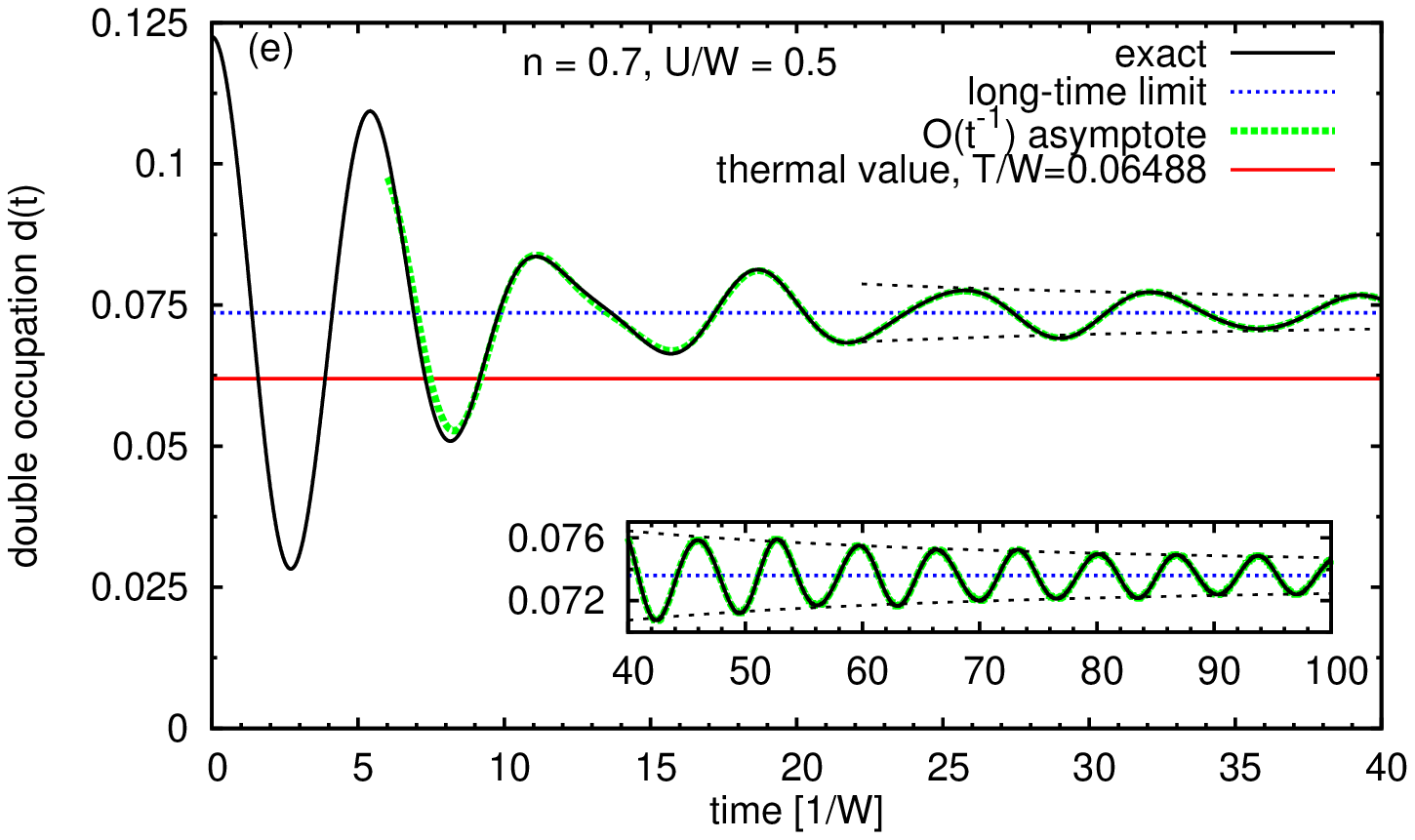}%
      \hspace*{\fill}%
    }
    \centerline{%
      \hspace*{\fill}%
      \includegraphics[width=0.48\hsize,clip=true]{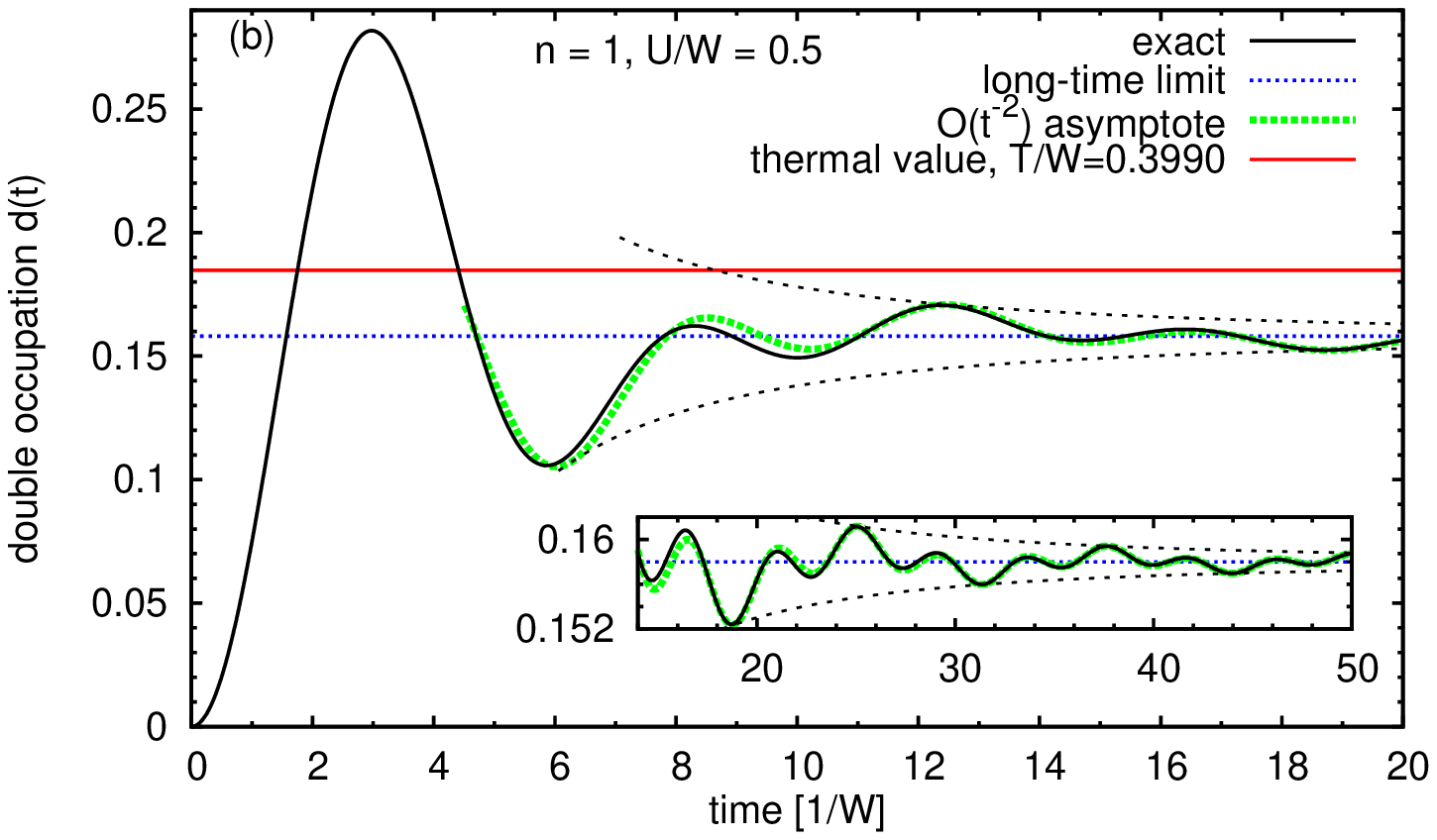}%
      \hspace*{\fill}%
      \includegraphics[width=0.48\hsize,clip=true]{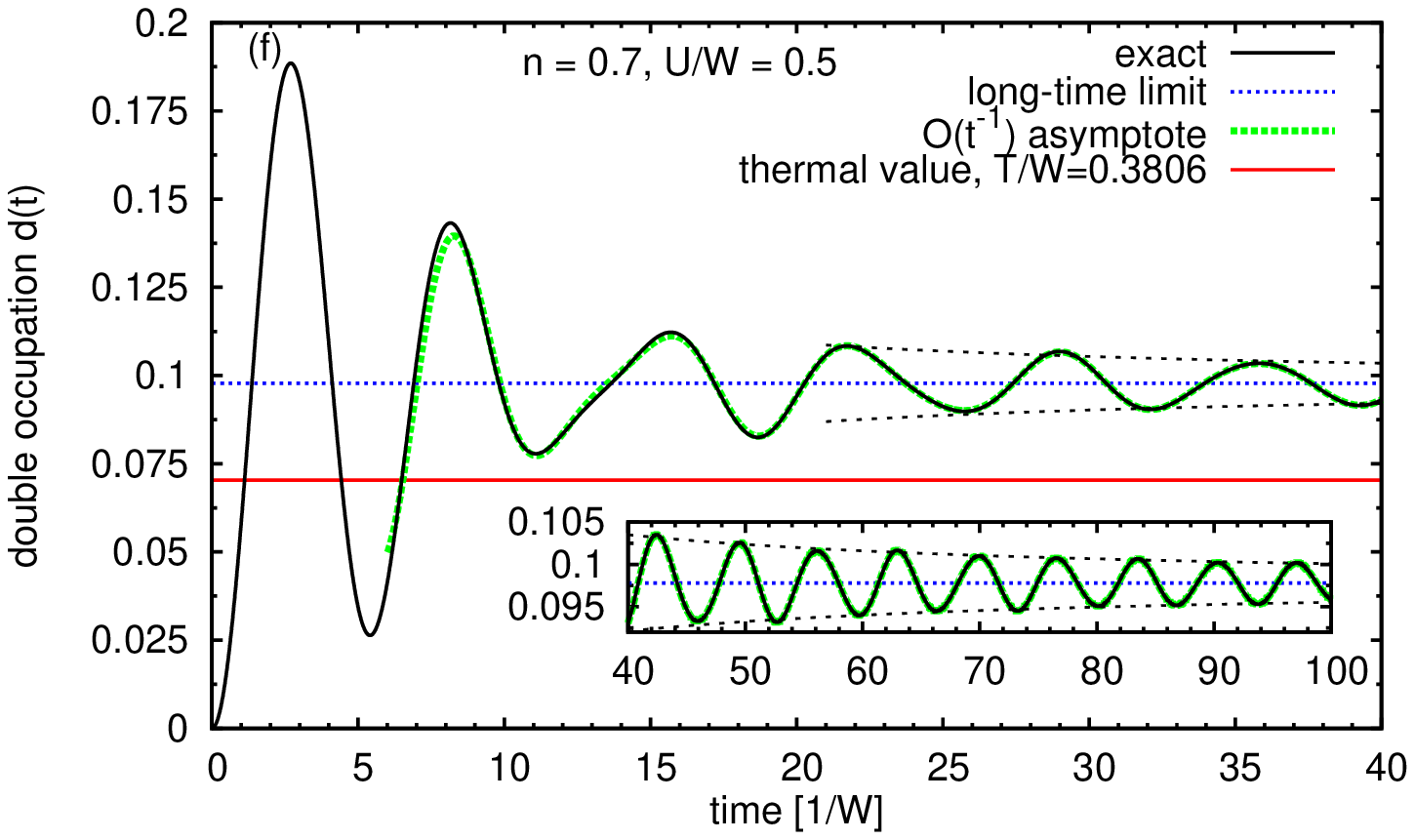}%
      \hspace*{\fill}%
    }
    \centerline{%
      \hspace*{\fill}%
      \includegraphics[width=0.48\hsize,clip=true]{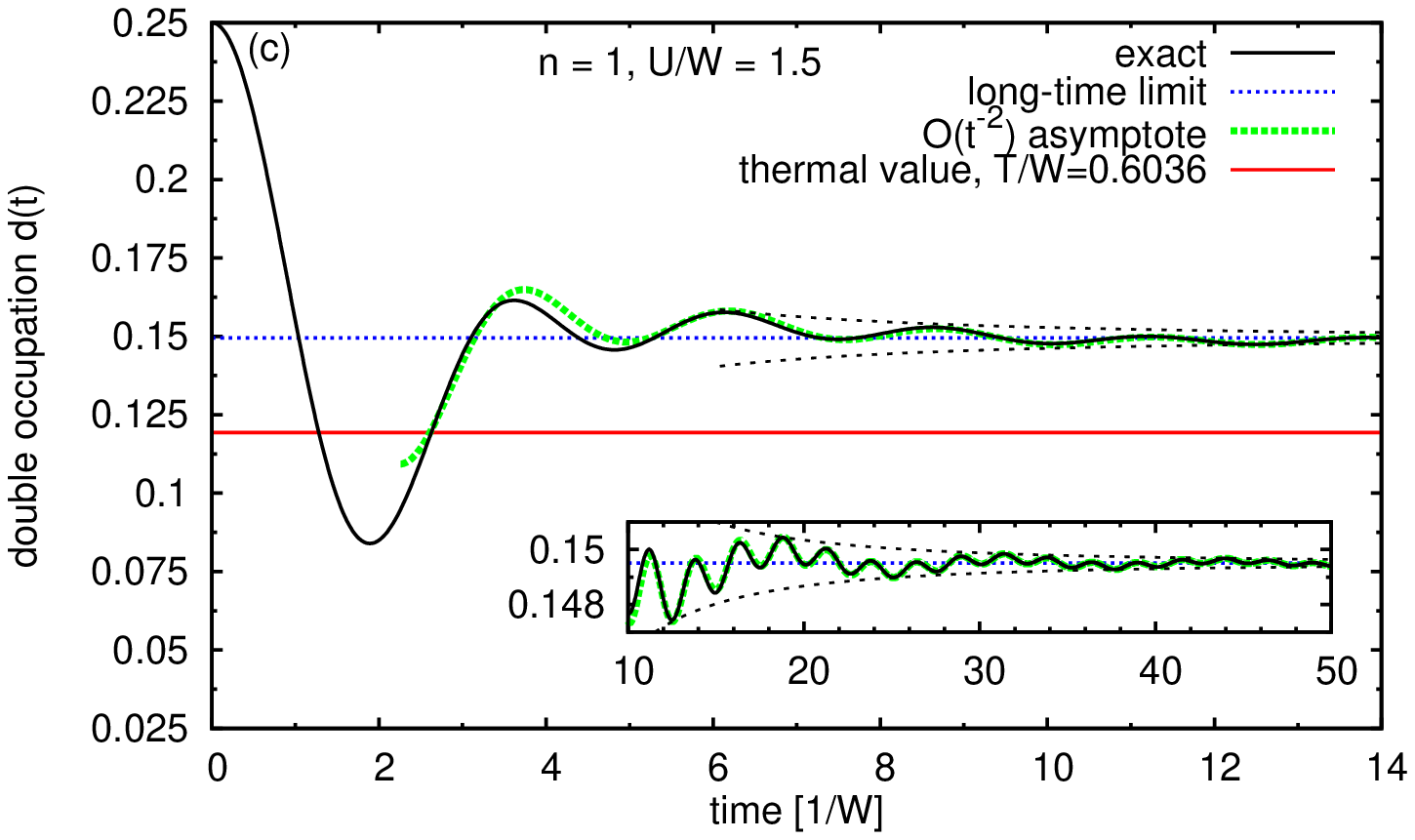}%
      \hspace*{\fill}%
      \includegraphics[width=0.48\hsize,clip=true]{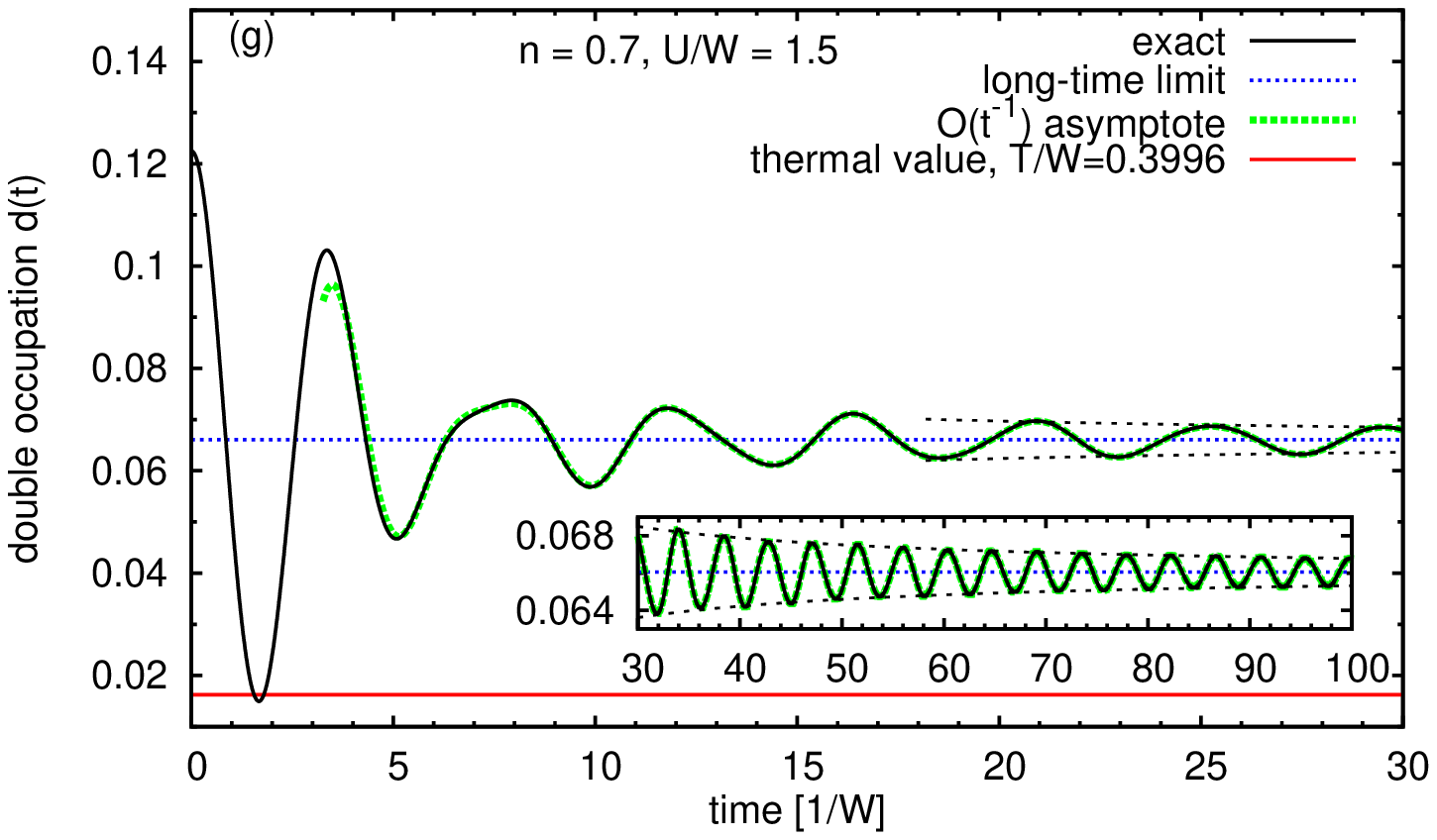}%
      \hspace*{\fill}%
    }
    \centerline{%
      \hspace*{\fill}%
      \includegraphics[width=0.48\hsize,clip=true]{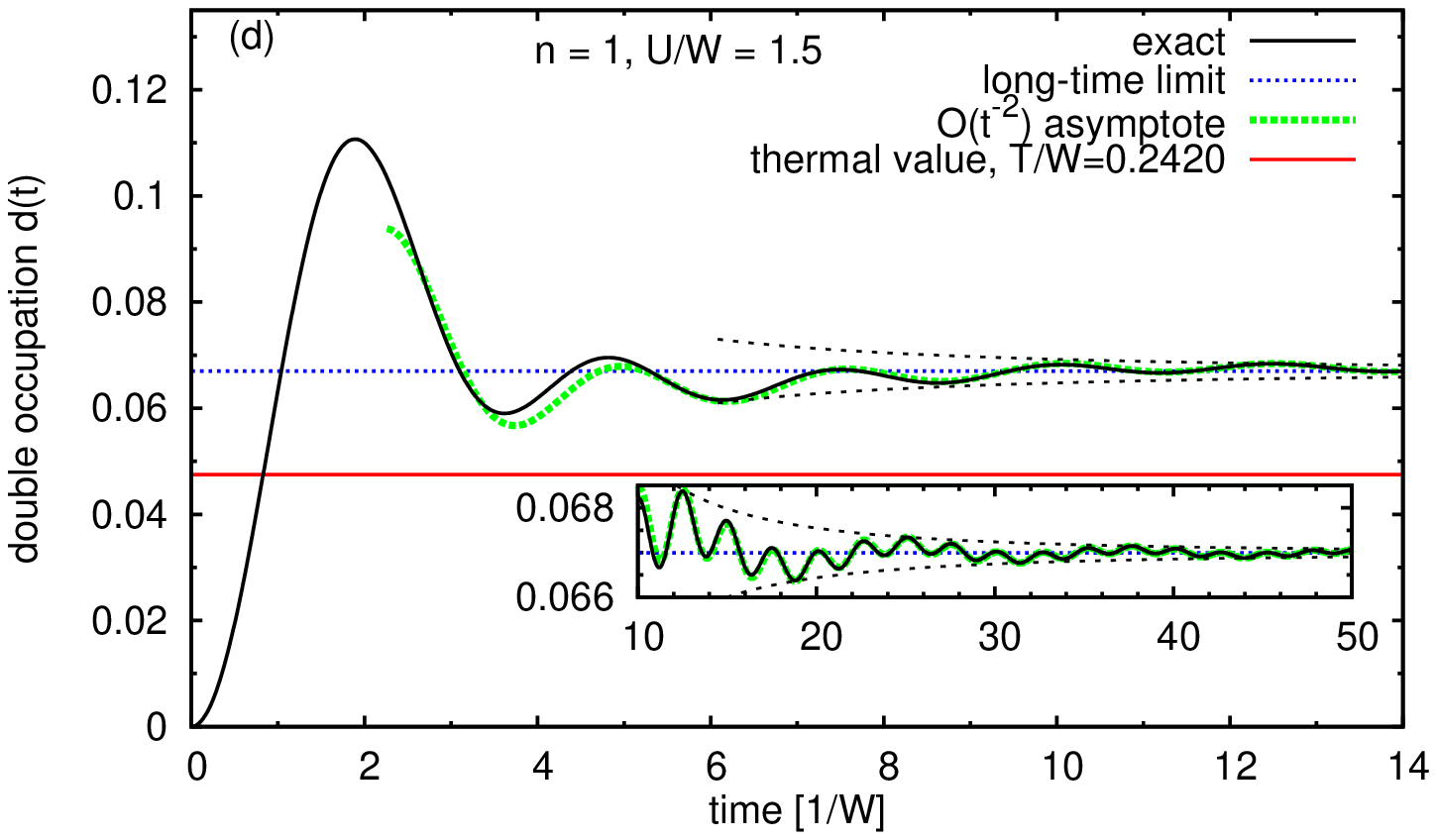}%
      \hspace*{\fill}%
      \includegraphics[width=0.48\hsize,clip=true]{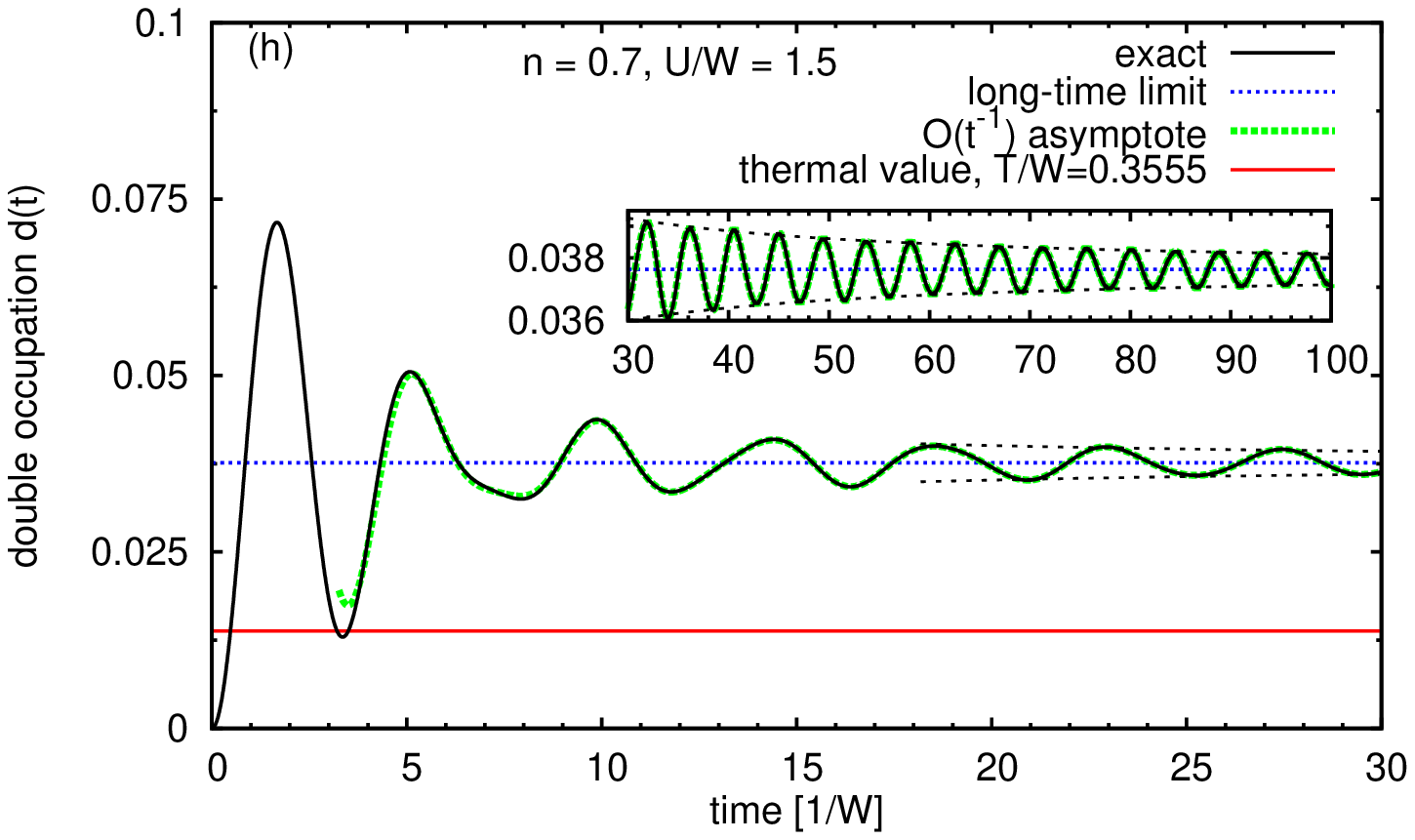}%
      \hspace*{\fill}%
    }
    \caption{(Color online)
      Relaxation of the double occupation $d(t)$ after an
      interaction quench to $U$ $=$ $0.5$ (top half) or $1.5$ (bottom
      half), for density $n$ $=$ $1$ (left column) and $n$ $=$ $0.7$
      (right column), from the metallic state (\ref{eq:inimetal}) in
      (a,c,e,g) and from the insulating state (\ref{eq:iniMott}) in
      (b,d,f,h).  The dashed black line shows the leading-order
      envelope of the asymptotic behavior.  The leading-order
      asymptote is marked by the dashed green line and the long-time
      limit by the dotted blue line. The solid red line marks the
      grand-canonical prediction for a temperature and chemical
      potential that corresponds to the internal energy and density of
      the final state.  The insets shows the large-$t$ behavior in
      more detail.\label{fig:docc}}
  \end{figure*}

  The results for $d(t)$ for several quenches are shown in
  Fig.~\ref{fig:docc}, %
  together with the long-time limit (dotted blue lines) and the
  thermodynamic prediction (solid red lines).  The latter is
  determined from the exact grand-canonical potential $f(T,\mu,U)$ $=$
  $-(T/L)$ $\TR[\exp(-(H-\mu
  N)/T)]$~\cite{Gebhard92,Gebhard94b,Gebhard97}, using the temperature
  $T$ and chemical potential $\mu$ that correspond to the same
  internal energy and density as the final state.  The density is
  given by $\expval{N}_{\text{gcan}}/L$ $=$ $\partial f/\partial\mu$,
  which yields the chemical potential $\mu(T,n,U)$ by inversion, and
  the internal energy per site is $e(T,\mu,U)$ $=$
  $\expval{H}_{\text{gcan}}/L$ $=$ $-T^2\partial (f/T)/\partial T$ $-$
  $\partial f/\partial\mu$. The temperature $T$ is obtained as the
  solution of $e(T,\mu(T,n,U),U)$ $=$ $\expval{H}_{t>0}$, where the
  energy in the final state is $\expval{H}_{t>0}$ $=$
  $-(2-n)n/4+U^2/4$ for the metallic state (\ref{eq:inimetal}) and
  $\expval{H}_{t>0}$ $=$ $-(1-n)n/4$ for the insulating state
  (\ref{eq:iniMott}).  Finally, the thermal value of the double
  occupation per site is obtained as $d(T,\mu,U)$ $=$
  $\expval{n_{i\uparrow}n_{i\downarrow}}_{\text{gcan}}$ $=$ $\partial
  f/\partial U$, evaluated at the determined temperature $T$ and
  chemical potential $\mu(T,n,U)$.

  The long-time limit of $d(t)$ clearly differs from the thermodynamic
  prediction (see Fig.~\ref{fig:docc}). This is the expected behavior
  for an integrable system with an infinite number of constants of
  motion.  They act as constraints on the accessible states and must
  be taken into account into the statistical description of the steady
  state.  As discussed in the Introduction, this can be achieved by
  employing a GGE~\cite{Rigol06+07}. For the $1/r$ Hubbard
  chain~(\ref{eq:GR-model}) the effective bosonic Hamiltonian is
  indeed of the form~(\ref{eq:effH}), where $\alpha$ labels the pairs
  of bosonic pseudomomenta (${\calK}$,${\cal K}+\Delta$) and the
  constants of motion $\calI_\alpha$ are the projectors onto one of
  the two eigenstates $\ket{\nu_{\calK,\calK+\Delta}}$ of
  $\bm{H}_{\calK}$ $\equiv$ $\bm{T}_{{\calK},{\calK}+\Delta{\calK}}$
  $+$ $U\bm{D}_{{\calK},{\calK}+\Delta{\calK}}$
  [Eq.~\ref{eq:GRM_Heff}].  For the initial
  states~(\ref{eq:initialstates}) of the form $\ket{\Psi(0)}$ $=$
  $\prod_{\calK}'\,\ket{\psi_{\calK,\calK+\Delta}}$ the statistical
  operator of the GGE~(\ref{eq:gge}) becomes%
  \begin{align}%
    \rho_{\text{G}}^\phdagger
    &=
    \prod_{\calK}{}'\,
    \sum_{\nu=1}^{2}
    \;
    |
    \braket%
    {\nu_{\calK,\calK+\Delta}}%
    {\psi_{\calK,\calK+\Delta}}
    |^2
    \;
    \ket{\nu_{\calK,\calK+\Delta}}
    \bra{\nu_{\calK,\calK+\Delta}}
    \,,\label{eq:GRM_gge}%
  \end{align}%
  from which we obtain
  $\expval{n_{i\uparrow}n_{i\downarrow}}_{\text{G}}^\phdagger$ as
  $c_+$ and $c_-/U$ for the two types of quenches.  The double
  occupation that is predicted by the GGE thus \emph{agrees precisely
    with the long-time limit [Eq.~(\ref{eq:longtimelimit})]}, i.e.,
  $\expval{n_{i\uparrow}n_{i\downarrow}}_{\text{G}}^\phdagger$ $=$
  $\lim_{t\to\infty}d(t)$.  In the next section we discuss general
  rules when GGEs are valid, which will explain in particular why the
  GGE gives the correct stationary value of the double occupation in
  the $1/r$ Hubbard model.


  \vspace*{3mm}

  \section{Validity of generalized Gibbs ensembles}\label{sec:gibbsensembles}

  \subsection*{Long-time average and diagonal ensemble}

  In order to address the validity of GGEs we first need to determine
  the long-time limit of a quantum system after an arbitrary quench.
  Suppose that an isolated system is prepared at time $t=0$ in an
  initial state which is described by the density matrix $\rho_0$,
  while the time evolution for $t>0$ is governed by an arbitrary
  time-independent Hamiltonian $H$. For an initial pure state (e.g.,
  as in Eq.~(\ref{eq:initialstates})) the density matrix $\rho_0$ has
  the form $\rho_0$ $=$ $\ket{\Psi(0)}\bra{\Psi(0)}$, whereas $\rho_0$
  $=$ $\sum_n p_n \ket{\Psi_n}\bra{\Psi_n}$ for a statistical mixture
  of orthogonal states $\ket{\Psi_n}$ with probabilities $p_n$ (e.g.,
  for an initial grand-canonical ensemble, as in
  Ref.~\onlinecite{Eckstein07}).

  For $t\geq0$ the time evolution of the density matrix is given by
  \begin{align}
    \rho(t)
    &=
    e^{iHt}
    \rho_0
    e^{-iHt}
    \,,\label{eq:rho-of-t}
  \end{align}
  and the expectation value of an observable $\OBS$ is
  \begin{align}
    \expval{\OBS}_t
    &=
    \TR{[\OBS\rho(t)]}
    \ifnarrow\nonumber\\\myamp\fi
    =
    \sum_{{nn'gg'}}
    e^{-i(E_n-E_{n'})t}
    \bra{ng}
    \OBS{}
    \ket{n'g'}
    \bra{n'g'}
    \rho_0
    \ket{ng}
    \,,
  \end{align}
  where $\ket{ng}$ are the eigenstates of $H$ with energies $E_n$, and
  $g$ labels possible degeneracies.  If the long-time limit
  $\lim_{t\to\infty}$ $\expval{\OBS}_t$ exists, then it is necessarily
  equal to the long-time average $\overline{\expval{\OBS}}$,
  \begin{align}
    \overline{\expval{\OBS}}
    &=
    \lim_{T\to\infty}
    \frac{1}{T}
    \int\limits_0^T
    \!dt\,
    \expval{\OBS}_t
    \ifnarrow\nonumber\\\myamp\fi
    =
    \sum_{ngg'}
    \bra{ng}
    \OBS{}
    \ket{ng'}
    \bra{ng'}
    \rho_0
    \ket{ng}
    \,,\label{eq:OBSaverage}
  \end{align}%
  assuming that the limit can be taken termwise (which is allowed for
  the large but finite systems that we have in mind).  In a steady
  state the system is thus described by the ``diagonal
  ensemble''~\cite{vonNeumann1929,Deutsch1991a,Brody2007,Rigol2008a},
  \begin{align}
    \rho_{\text{diag}}^\phdagger
    &=
    \sum_{ngg'}
    \ket{ng}
    \bra{ng}
    \rho_0
    \ket{ng'}
    \bra{ng'}
    =
    \sum_n P_n\,\rho_0\,P_n
    \,.\label{eq:rho_diag_def}
  \end{align}
  Here $P_n$ $=$ $\sum_{g}\ket{ng}\bra{ng}$ is the projector onto the
  subspace spanned by the eigenvectors corresponding to the energy
  eigenvalue $E_n$.  The statistical operator
  $\rho_{\text{diag}}^\phdagger$ correctly describes the long-time
  limit, if it exists, of any observable $\OBS$, i.e.,
  $\overline{\expval{\OBS}}$ $=$
  $\TR{[\OBS\rho_{\text{diag}}^\phdagger]}$.

  The diagonal ensemble correctly yields any stationary expectation
  value, regardless of the transient behavior.  However, not a lot is
  gained by regarding $\rho_{\text{diag}}^\phdagger$ as a
  ``statistical'' prediction for the steady state, because each energy
  eigenstate contributes according to the initial conditions given by
  $\rho_0$.  For a nonintegrable system one expects, by contrast,
  that only a few conserved quantities such as energy and particle number
  need to be fixed for a successful statistical description in terms
  of thermal Gibbs ensembles, which can emerge from the diagonal
  ensemble by means of ``eigenstate
  thermalization''~\cite{Deutsch1991a,Srednicki1994a,Rigol2008a}.
  Similarly one can ask when a GGE~(\ref{eq:gge}) for an integrable
  system yields the same prediction as the diagonal
  ensemble~(\ref{eq:rho_diag_def}).  This is discussed in the next
  subsection.

  \vspace*{1mm}

  \subsection*{Gibbs ensemble for an integrable system}

  \vspace*{1mm}

  We now consider an integrable system whose time evolution after the
  quench is governed by the effective Hamiltonian~(\ref{eq:effH}).
  For simplicity we consider two typical cases of Hamiltonians only.
  Either~{\mycasea} the constants of motion $\calI_\alpha$ have the
  eigenvalues 0 and 1 and can thus be represented by fermions or
  hard-core bosons, $\calI_\alpha$ $=$ $a_{\alpha}^\mydagger
  a_{\alpha}^\phdagger$, with
  $[a_{\alpha}^\phdagger,a_{\beta}^\mydagger]_\pm$ $=$
  $\delta_{\alpha\beta}$, $(a_{\alpha}^\phdagger)^2$ $=$
  $(a_{\alpha}^\mydagger)^2$ $=$ $0$; or~{\mycaseb} the $\calI_\alpha$
  have the eigenvalues $0,1,2\ldots$ and can be represented by bosons,
  $\calI_\alpha$ $=$ $b_{\alpha}^\mydagger b_{\alpha}^\phdagger$, with
  $[b_{\alpha}^\phdagger,b_{\beta}^\mydagger]$ $=$
  $\delta_{\alpha\beta}$.  Examples for case~{\mycasea} are the
  effective Hamiltonians for hard-core bosons in one
  dimension~\cite{Rigol06+07}, free fermions with quenched
  disorder~\cite{Eckstein07}, and the $1/r$ fermionic Hubbard chain
  (Sec.~\ref{sec:hubbardmodel}), whereas case~{\mycaseb} applies to
  the Luttinger model~\cite{Cazalilla06}.  For both cases the Lagrange
  multipliers $\lambda_\alpha$ in Eq.~(\ref{eq:gge}) are then given
  by~{\mycasea} $\ln[\expval{\calI_\alpha}_0^{-1}-1]$ and ~{\mycaseb}
  $\ln[\expval{\calI_\alpha}_0^{-1}+1]$.

  The Hamiltonian~(\ref{eq:effH}) has the eigenstates $\ket{\bm{m}}$
  with occupation numbers $\calI_\alpha\ket{\bm{m}}$ $=$
  $m_\alpha\ket{\bm{m}}$ and energy eigenvalues $E_{\bm{m}}$ $=$
  $\sum_\alpha\epsilon_\alpha m_\alpha$. For simplicity \emph{we
    assume that the degeneracy of energy eigenvalues is irrelevant},
  i.e., the observable $\OBS$ or the initial-state density matrix
  $\rho_0$ are diagonal in the subspace of eigenvectors $\ket{\bm{m}}$
  with the same energy. This assumption will be examined in detail at
  the end of this section. {}From Eq.~(\ref{eq:OBSaverage}) the
  diagonal ensemble and long-time average are then given
  by
  \begin{align}%
    \overline{\expval{\OBS}}
    &=
    \sum_{\bm{m}}
    \bra{\bm{m}}
    \OBS{}
    \ket{\bm{m}}
    \bra{\bm{m}}
    \rho_0
    \ket{\bm{m}}
    \,.\label{eq:OBSaverage_diag}
  \end{align}%
  Is this the steady-state value predicted by the GGE~(\ref{eq:gge})?
  We answer this question for two types of observables: for
  case~{\mycasea} we consider the observable
  \begin{align}
    A=\sum_{\substack{\alpha_1\cdots\alpha_m\\\beta_1\cdots\beta_m}}\;
    A^{\alpha_1\cdots\alpha_m}_{\beta_1\cdots\beta_m}\;
    a_{\alpha_1}^\mydagger\cdots
    a_{\alpha_m}^\mydagger
    a_{\beta_m}^\phdagger\cdots
    a_{\beta_1}^\phdagger
    \,,\label{eq:Aobservable}
  \end{align}
  while for case~{\mycaseb} we allow for powers of the bosonic
  operators and consider (for $r_i,s_j\geq1$)
\ifnarrow
  \begin{multline}
    B=\sum_{\substack{\alpha_1\cdots\alpha_m,r_1\cdots r_m\\\beta_1\cdots\beta_m,s_1\cdots s_m}}\;
    B^{\alpha_1\cdots\alpha_m,r_1\cdots r_m}_{\beta_1\cdots\beta_m,s_1\cdots s_m}\;
    \\
    (b_{\alpha_1}^\mydagger)^{r_1}\cdots
    (b_{\alpha_m}^\mydagger)^{r_m}
    (b_{\beta_m}^\phdagger)^{s_m}\cdots
    (b_{\beta_1}^\phdagger)^{s_1}
    \,,\label{eq:Bobservable}
  \end{multline}
\else
  \begin{align}
    B=\sum_{\substack{\alpha_1\cdots\alpha_m,r_1\cdots r_m\\\beta_1\cdots\beta_m,s_1\cdots s_m}}\;
    B^{\alpha_1\cdots\alpha_m,r_1\cdots r_m}_{\beta_1\cdots\beta_m,s_1\cdots s_m}\;
    (b_{\alpha_1}^\mydagger)^{r_1}\cdots
    (b_{\alpha_m}^\mydagger)^{r_m}
    (b_{\beta_m}^\phdagger)^{s_m}\cdots
    (b_{\beta_1}^\phdagger)^{s_1}
    \,,\label{eq:Bobservable}
  \end{align}
\fi
  We assume without loss of generality that
  $B^{\alpha_1\cdots\alpha_m,r_1\cdots
    r_m}_{\beta_1\cdots\beta_m,s_1\cdots s_m}$ vanishes whenever two
  indices $\alpha_i$ or two indices $\beta_j$ are the same.

  It is straightforward to obtain the long-time
  average~(\ref{eq:OBSaverage_diag}) and GGE average~(\ref{eq:gge}) of
  the observables $A$ and $B$ by using the occupation number basis
  $\ket{\bm{m}}$ and the fixed GGE averages
  $\expval{\calI_\alpha}_\text{G}^\phdagger$ $=$
  $\expval{\calI_\alpha}_{0}^\phdagger$.  In case~{\mycasea} we find
  \begin{subequations}%
    \begin{align}%
      \overline{\expval{A}}
      &=
      \sum_{\alpha_1\cdots\alpha_m}\;
      \widetilde{A}_{\alpha_1\cdots\alpha_m} 
      \,\left\langle{\,\textstyle\prod\limits_{i=1}^m\calI_{\alpha_i}}\right\rangle_{\!\!0}^\phdagger
      \,,\label{eq:Atimeaverage}
      \\
      \expval{A}_{\text{G}}^\phdagger
      &=
      \sum_{\alpha_1\cdots\alpha_m}\;
      \widetilde{A}_{\alpha_1\cdots\alpha_m} 
      \,{\textstyle\prod\limits_{i=1}^m\expval{\,\calI_{\alpha_i}}_0^\phdagger}
      \,,\label{eq:AGGEaverage}
    \end{align}\label{eq:Aaverages}%
  \end{subequations}%
  where we used the identity
  \begin{align}    
    \left\langle{\,\textstyle\prod\limits_{i=1}^m\calI_{\alpha_i}}\right\rangle_{\!\!\text{G}}^\phdagger
    =
    {\textstyle\prod\limits_{i=1}^m\expval{\,\calI_{\alpha_i}}_\text{G}^\phdagger}
    =
    {\textstyle\prod\limits_{i=1}^m\expval{\,\calI_{\alpha_i}}_0^\phdagger}
    \label{eq:Aidentity}
  \end{align}%
  in the second line. In case~{\mycaseb} we have
  \begin{subequations}%
    \begin{align}%
      \overline{\expval{B}}
      &=
      \sum_{\alpha_1\cdots\alpha_m}\;
      \widetilde{B}_{\alpha_1\cdots\alpha_m}^{r_1\cdots r_m}
      \,\left\langle{\,\textstyle\prod\limits_{i=1}^m
          (b_{\alpha_i}^\mydagger)^{r_i}(b_{\alpha_i}^\phdagger)^{r_i}
        }\right\rangle_{\!\!0}^\phdagger
      \nonumber\\
      &=
      \sum_{\alpha_1\cdots\alpha_m}\;
      \widetilde{B}_{\alpha_1\cdots\alpha_m}^{r_1\cdots r_m} 
      \,\left\langle{\,\textstyle\prod\limits_{i=1}^m
          \prod\limits_{k=0}^{r_i-1}(\calI_{\alpha_i}-k)
        }\right\rangle_{\!\!0}
      \,,\label{eq:Btimeaverage}
      \\
      \expval{B}_{\text{G}}^\phdagger
      &=
      \sum_{\alpha_1\cdots\alpha_m}\;
      \widetilde{B}_{\alpha_1\cdots\alpha_m}^{r_1\cdots r_m} 
      \,\left\langle{\,\textstyle\prod\limits_{i=1}^m
          (b_{\alpha_i}^\mydagger)^{r_i}(b_{\alpha_i}^\phdagger)^{r_i}
        }\right\rangle_{\!\!\text{G}}^\phdagger
      \nonumber\\
      &=
      \sum_{\alpha_1\cdots\alpha_m}\;
      \widetilde{B}_{\alpha_1\cdots\alpha_m}^{r_1\cdots r_m}
      \,{\textstyle\prod\limits_{i=1}^m\left[\,r_i!\,(\expval{\calI_{\alpha_i}}_0^\phdagger)^{r_i}\,\right]}
      \,,\label{eq:BGGEaverage}
    \end{align}%
    \label{eq:Baverages}%
  \end{subequations}%
  where we used the bosonic operator identity
  \begin{align}
    {\textstyle
      (b_{\alpha_i}^\mydagger)^{r_i}
      (b_{\alpha_i}^\phdagger)^{r_i}
    }
    =
    {\textstyle
      \prod\limits_{k=0}^{r_i-1}(\calI_{\alpha_i}-k)
    }
  \end{align}
  in the first line, and the identity
\ifnarrow
  \begin{align}    
    \bigg\langle{\,\textstyle\prod\limits_{i=1}^m}&{\textstyle
        (b_{\alpha_i}^\mydagger)^{r_i}(b_{\alpha_i}^\phdagger)^{r_i}
      }\bigg\rangle_{\!\!\text{G}}^\phdagger
    =
    {\textstyle\prod\limits_{i=1}^m\expval{(b_{\alpha_i}^\mydagger)^{r_i}(b_{\alpha_i}^\phdagger)^{r_i}}_\text{G}^\phdagger}
    \nonumber\\
    &=
    {\textstyle\prod\limits_{i=1}^m\left\langle\,
        \prod\limits_{k=0}^{r_i-1}(\calI_{\alpha_i}-k)
      \right\rangle_{\!\!\text{G}}^\phdagger}
    =
    {\textstyle\prod\limits_{i=1}^m\left[\,r_i!\,(\expval{\calI_{\alpha_i}}_\text{G}^\phdagger)^{r_i}\,\right]}
    \nonumber\\
    &=
    {\textstyle\prod\limits_{i=1}^m\left[\,r_i!\,(\expval{\calI_{\alpha_i}}_0^\phdagger)^{r_i}\,\right]}
    \label{eq:Bidentity}
  \end{align}%
\else
  \begin{align}    
    \bigg\langle{\,\textstyle\prod\limits_{i=1}^m}&{\textstyle
        (b_{\alpha_i}^\mydagger)^{r_i}(b_{\alpha_i}^\phdagger)^{r_i}
      }\bigg\rangle_{\!\!\text{G}}^\phdagger
    =
    {\textstyle\prod\limits_{i=1}^m\expval{(b_{\alpha_i}^\mydagger)^{r_i}(b_{\alpha_i}^\phdagger)^{r_i}}_\text{G}^\phdagger}
    =
    {\textstyle\prod\limits_{i=1}^m\left\langle\,
        \prod\limits_{k=0}^{r_i-1}(\calI_{\alpha_i}-k)
      \right\rangle_{\!\!\text{G}}^\phdagger}
    \nonumber\\
    &=
    {\textstyle\prod\limits_{i=1}^m\left[\,r_i!\,(\expval{\calI_{\alpha_i}}_\text{G}^\phdagger)^{r_i}\,\right]}
    =
    {\textstyle\prod\limits_{i=1}^m\left[\,r_i!\,(\expval{\calI_{\alpha_i}}_0^\phdagger)^{r_i}\,\right]}
    \label{eq:Bidentity}
  \end{align}%
\fi
  in the second line.  Furthermore we defined the
  permu\-tation-aver\-aged matrix elements
  $\widetilde{A}_{\alpha_1\cdots\alpha_m}$ $=$ $\sum_P$ $(\mp1)^P$
  $A^{\alpha_1\cdots\alpha_m}_{\alpha_{P1}\cdots\alpha_{Pm}}$ and
  $\widetilde{B}_{\alpha_1\cdots\alpha_m}^{r_1\cdots r_m} $ $=$ $\sum_P$
  $B^{\alpha_1\cdots\alpha_m,r_1\cdots
    r_m}_{\alpha_{P1}\cdots\alpha_{Pm},r_{P1}\cdots r_{Pm}}$.

  {}From these results we obtain rather general sufficient conditions
  for the validity of the GGE predictions, namely the
  \emph{factorization of initial-state expectation values}
  of~{\mycasea} products or~{\mycaseb} polynomials of the constants of
  motion~$\calI_\alpha$ as follows:
  \begin{subequations}%
    \begin{align}%
      &\text{If~}
      \left\langle{\,\textstyle\prod\limits_{i=1}^m\calI_{\alpha_i}}\right\rangle_{\!\!0}^\phdagger
      =
      {\textstyle\prod\limits_{i=1}^m\expval{\,\calI_{\alpha_i}}_0^\phdagger}
      &&\text{then~~~}
      \overline{\expval{A}}
      =\expval{A}_{\text{G}}^\phdagger
      \,.\label{eq:factorizationA}%
      \\
      &\text{If~}
      \left\langle{\,\textstyle\prod\limits_{i=1}^m
          \prod\limits_{k=0}^{r_i-1}(\calI_{\alpha_i}-k)
        }\right\rangle_{\!\!0}
      =
      \ifnarrow
      \makebox[0pt][l]{$%
        \textstyle\prod\limits_{i=1}^m\left[\,r_i!\,(\expval{\calI_{\alpha_i}}_0^\phdagger)^{r_i}\,\right]%
      $}
      \else
        \textstyle\prod\limits_{i=1}^m\left[\,r_i!\,(\expval{\calI_{\alpha_i}}_0^\phdagger)^{r_i}\,\right]
      \fi
      \ifnarrow\nonumber\\\myamp\fi
      &&\text{then~~~}
      \overline{\expval{B}}
      =\expval{B}_{\text{G}}^\phdagger
      \,.\,\label{eq:factorizationB}%
    \end{align}\label{eq:factorizations}%
  \end{subequations}%
  The $\alpha_i$ and $r_i$ in (\ref{eq:factorizationA}) and
  (\ref{eq:factorizationB}) are those for which
  $\widetilde{A}_{\alpha_1\cdots\alpha_m}$ and
  $\widetilde{B}_{\alpha_1\cdots\alpha_m}^{r_1\cdots r_m}$,
  respectively, are nonzero. The criteria (\ref{eq:factorizations})
  are the central result of this section, and we now discuss their
  implications in detail.

  First we note that for simple observables, which involve at most one
  factor $\calI_\alpha$, the factorizations (\ref{eq:factorizations})
  occur trivially.  This is the case, e.g., for the double occupancy
  in the $1/r$ Hubbard model, explaining why the
  GGE~(\ref{eq:GRM_gge}) works in this case.  Typical observables of
  an \emph{interacting} integrable system, however, are often rather
  complicated when expressed in terms of the constants of motion
  $\calI_{\alpha}$.  Thus all correlations between the constants of
  motion must vanish in the initial state in order for
  (\ref{eq:factorizations}) to be fulfilled.  This seemingly
  restrictive condition can nevertheless be met, because often the
  initial state is not strongly correlated in terms of the
  $\calI_\alpha$.  Moreover, some correlations among the
  $\calI_\alpha$ are allowed provided that their contribution to the
  sum (\ref{eq:Aaverages}) or (\ref{eq:Baverages}) is negligible,
  e.g., if it vanishes in the thermodynamic limit.

  For example, one-dimensional hard-core bosons are represented by a
  free-fermion Hamiltonian.  For an alternating potential (as studied
  in Refs.~\onlinecite{Rigol06+07}) the initial state contains
  correlations only between the fermionic momentum number operators
  $n_k$ and $n_{k+\pi}$.  One can then show from (\ref{eq:Aaverages})
  that the GGE~(\ref{eq:gge}) makes correct predictions, up to
  finite-size corrections, for observables that are restricted to a
  finite region of real space.  Another example is the fermionic
  Luttinger model, which maps to a free-boson Hamiltonian.  For an
  interaction quench (studied in Ref.~\onlinecite{Cazalilla06}) the
  initial state is a product state with correlations only between the
  bosonic momentum occupation $n_q$ and $n_{-q}$.

  On the other hand, correlations between constants of motion in the
  initial state, which remain for all times, cannot be described by
  the GGE~(\ref{eq:gge}), as noted in Ref.~\onlinecite{Gangardt2008}.
  However, in interacting integrable systems, such observables usually
  correspond to complicated many-particle operators in terms of the
  original microscopic degrees of freedom, and thus are not measurable
  in practice. As mentioned in the Introduction, the microcanonical
  Gibbs ensemble faces the same problem: one can always construct
  fine-grained observables that do not thermalize.

  Can a GGE be improved if it does not yield the correct long-time
  average? The minimal necessary extension of the ensemble depends on
  the observable in question, but as one sees from
  Eq.~(\ref{eq:factorizations}), it always suffices to
  fix not only the expectation values of the constants of motion
  $\calI_\alpha$ but also the expectation values of \emph{all of
  their products}, i.e., by using
  \begin{align}
    \widetilde{\rho}_{\text{G}}^\phdagger
    &\propto
    \exp\left(
      -\sum_\alpha\lambda_\alpha\calI_\alpha
      -\sum_{ab}\lambda_{ab}\calI_\alpha\calI_\beta
      -\cdots
    \right)
    \,,\label{eq:rhotilde}%
  \end{align}
  where the Lagrange multipliers are chosen to fix \emph{all}
  products,
  $\TR[\calI_{a_1}\cdots\calI_{a_m}\widetilde{\rho}_{\text{G}}^\phdagger]$
  $=$ $\TR[\calI_{a_1}\cdots\calI_{a_m}\rho_{0}^\phdagger]$.  Then it
  follows immediately that
  $\TR[A\widetilde{\rho}_{\text{G}}^\phdagger]$ $=$
  $\overline{\expval{A}}$ and
  $\TR[B\widetilde{\rho}_{\text{G}}^\phdagger]$ $=$
  $\overline{\expval{B}}$ for the observables considered above.
  Thus \emph{any steady state can be described by a
    sufficiently extended GGE}, i.e., by fixing sufficiently many
  products of constants of motion.  

  While fixing all products as in Eq.~(\ref{eq:rhotilde}) yields the
  exact long-time average, this extension of the GGE can hardly be
  regarded as a statistical description of the steady state (as noted
  in the previous subsection for the diagonal ensemble), because it
  uses almost the full information about the initial state.  In fact,
  any nondegenerate Hamiltonian $H$ acting on a Hilbert space of
  dimension $h$ has $h-1$ pairwise commuting and linearly independent
  constants of motion~\cite{Rigol2008a}; fixing all of them in a GGE
  recovers the diagonal ensemble~\cite{Brody2007}.  For a
  nonintegrable system one can choose, e.g., the projectors onto the
  eigenstates of $H$~\cite{Brody2007}, or linearly independent integer
  powers of $H$~\cite{Manmana07}.  In practice, however, these
  extensions of the GGE are as hard to calculate as the long-time
  average.
  
  \vspace*{2mm}

  \subsection*{Degenerate energy levels}

  \vspace*{2mm}

  In the previous subsection we assumed that the degeneracy of energy
  levels is irrelevant [as defined above
  Eq.~(\ref{eq:OBSaverage_diag})], which allowed us to move from
  Eqs.~(\ref{eq:OBSaverage}-\ref{eq:rho_diag_def})
  to~(\ref{eq:OBSaverage_diag}).  Below we provide an example for
  which this assumption does not hold. In that case the
  expression~(\ref{eq:OBSaverage_diag}) for the long-time average
  cannot be used, and thus neither~(\ref{eq:Atimeaverage})
  nor~(\ref{eq:Btimeaverage}) are available.

  We consider a quench to $U$ $=$ $0$ in a general fermionic Hubbard
  model.  Fermions (with spin $\sigma$ $=$ $\uparrow,\downarrow$) on a
  Bravais lattice (with $L$ lattice sites) are prepared in a
  correlated unpolarized initial state $\rho_0$ with fixed densities
  $n_\uparrow$ $=$ $n_\downarrow$ $=$ $n/2.$ The time evolution is
  governed by the free Hamiltonian
  \begin{align}%
    H
    &=
    \sum_{\substack{ij\sigma}}
    V_{ij}
    c_{i\sigma}^\mydagger
    c_{j\sigma}^\phdagger
    =
    \sum_{\substack{{{k}}\sigma}}
    \epsilon_{{{k}}}\,
    c_{{{k}}\sigma}^\mydagger
    c_{{{k}}\sigma}^\phdagger
    \,,\label{eq:freefermions}%
  \end{align}%
  where ${{k}}$ labels the crystal momentum (we suppress the vector
  notation); periodic boundary conditions are assumed for simplicity.
  This Hamiltonian of the form~(\ref{eq:effH}), with the number
  operators $n_{{{k}}\sigma}$ $=$ $c_{{{k}}\sigma}^\mydagger
  c_{{{k}}\sigma}^\phdagger$ playing the role of the constants of
  motion $\calI_\alpha$. We are interested in the steady-state
  expectation value of the double occupation
  $n_{i\uparrow}n_{i\downarrow}$.

  Assuming again that the degeneracy of energy levels is irrelevant,
  we obtain for the long-time average~(\ref{eq:OBSaverage_diag}),
  using the basis $\ket{\bm{m}}$ $=$
  $\prod_{k\sigma}(c_{{{k}}\sigma}^\mydagger)^{m_{{{k}}\sigma}}\ket{0}$,
  \begin{align}
    \overline{\expval{n_{i\uparrow}n_{i\downarrow}}}
    &=
    \sum_{\bm{m}}
    \bra{\bm{m}}
    \rho_0
    \ket{\bm{m}}
    \,\frac{1}{L^2}
    \sum_{{{k}}{{k'}}}m_{{{k}}\uparrow}m_{{{k'}}\downarrow}
    \ifnarrow\nonumber\\\myamp\fi
    =
    \TR\Bigg[\frac{\rho_0}{L^2}
    \sum_{ij} n_{i\uparrow}n_{j\downarrow}
    \Bigg]
    =
    n_{\uparrow}n_{\downarrow}
    =
    \frac{n^2}{4}
    \,.\label{eq:zeroquench_timeaverage}
  \end{align}
  The same value is obtained from the canonical and grand-canonical
  ensemble, and also from the generalized Gibbs ensemble which uses
  the number operators $n_{{{k}}\sigma}$ as constants of motion:
  \begin{align}
    \expval{n_{i\uparrow}n_{i\downarrow}}_{\text{G}}^\phdagger
    &=
    \frac{1}{L^2}
    \sum_{{{k}}{{k'}}}
    \expval{n_{k\uparrow}}_{\text{G}}^\phdagger
    \expval{n_{k'\downarrow}}_{\text{G}}^\phdagger
    \ifnarrow\nonumber\\\myamp\fi
    =
    \frac{1}{L^2}
    \sum_{{{k}}{{k'}}}
    \expval{n_{k\uparrow}}_{\text{0}}^\phdagger
    \expval{n_{k'\downarrow}}_{\text{0}}^\phdagger
    =
    n_{\uparrow}n_{\downarrow}
    =
    \frac{n^2}{4}
    \,.\label{eq:zeroquench_GGE}
  \end{align}
  Thus we conclude that \emph{the double occupation thermalizes to the
    value $n^2/4$ after a quench to $U$ $=$ $0$ in any Hubbard model},
  provided that the degeneracy of energy levels is indeed irrelevant.

  Interestingly, this statement disagrees with our exact results for
  the $1/r$ Hubbard chain from Sec.~\ref{sec:hubbardmodel}: when
  quenching from $U$ $=$ $\infty$ to $0$ we obtained the long-time
  limit as $d_\infty$ $=$
  $n^2(3-2n)/6$~[Eq.~(\ref{eq:dinfty_results})].  This differs from
  the long-time average~(\ref{eq:zeroquench_timeaverage}) because the
  degeneracy of energy levels is in fact relevant for this quench: the
  initial-state density matrix does not factorize in the free-fermion
  basis and the linear dispersion $\epsilon_{{{k}}}$ $=$ $tk$ leads to
  a massive degeneracy for the free-fermion energy eigenstates
  $\ket{\bm{m}}$. Therefore the long-time limit is not given by
  Eq.~(\ref{eq:OBSaverage_diag}) and does not equal $n^2/4$.

  Furthermore this example shows that GGEs based on different
  representations of the constants of motion can yield different
  results.  The free-fermion GGE~(\ref{eq:zeroquench_GGE}) predicts
  the wrong value $n^2/4$, whereas the GGE~(\ref{eq:GRM_gge}) which
  uses the effective bosonic representation gives the correct value
  $n^2(3-2n)/6$. In fact the choice of constants of motion for the
  construction of a GGE~(\ref{eq:gge}) is always ambiguous, as
  discussed in the Introduction. Nevertheless it is possible to
  determine the correct GGE \emph{a priori}, i.e., without knowing the
  real-time dynamics, by verifying that the degeneracy of energy
  levels is irrelevant and that the
  conditions~(\ref{eq:factorizations}) are fulfilled.


  \bigskip
  
  \section{Conclusion}
  
  The exact real-time dynamics of the double occupation in the
  fermionic $1/r$ Hubbard chain shows that the presence of a Mott gap
  does not inhibit the relaxation after an interaction quench.  Its
  steady-state properties are correctly predicted by generalized Gibbs
  ensembles.  Furthermore we showed for a general class of integrable
  quantum systems that the GGE prediction equals the long-time
  average, provided that the observables or initial states are 
  sufficiently uncorrelated in terms of the constants of motion.

  \section*{Acknowledgements} 

  Useful discussions with Marcos Rigol, Stefan Kehrein, Corinna
  Kollath, Krzysztof Byczuk, and Dieter Vollhardt are gratefully
  acknowledged. This work was supported in part by the SFB 484 of the
  Deutsche Forschungsgemeinschaft.


\end{document}